\documentstyle[11pt,moriond,epsfig]{article}

\def\be{\begin{equation}}
\def\ee{\end{equation}}
\def\bea{\begin{eqnarray}}
\def\eea{\end{eqnarray}}

\def\dps{\displaystyle}
\begin{document}
\vspace*{4cm}
\title{End-to-end algorithm for hierarchical area searches for long-duration
GW
       sources for GEO 600}

\author{Bernard F. Schutz, M.Alessandra Papa }

\address{Albert Einstein Institut, Potsdam, Germany}

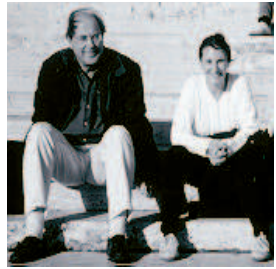
\maketitle\abstracts{We describe a hierarchical, highly parallel
computer algorithm to perform searches for
unknown sources of continuous gravitational waves --- spinning neutron
stars in the Galaxy --- over wide areas of the sky and wide frequency
bandwidths. We optimize the algorithm for an observing period of
4 months and an available computing power of 20~Gflops, in a
search for neutron stars resembling millisecond pulsars.  We show
that, if we restrict the search to the galactic plane, the method
will detect any star whose signal is stronger than 15 times the
$1\sigma$ noise level of a detector over that search period.  Since
on grounds of confidence the minimum identifiable signal should
be about 10 times noise, our algorithm does only 50\% worse than this
and runs on a computer with achievable processing speed.}

\section{Surveys for neutron stars}
In this paper we describe progress on developing an efficient computer
algorithm to search large areas of the sky for continuous gravitational wave
signals from previously unknown sources, most likely spinning neutron
stars.  The enormous computational cost of processing several months of
data by repeatedly applying matched filtering to the data for possible
sources lying in each resolved area of the sky is well
known.\cite{SCHUTZ1991a,Brady1998}  We have previously presented several
components of an algorithm that makes use of hierarchical methods to
improve search speeds and retain good
sensitivity.\cite{Schutz1998,gwda3,lisa}
Here we put the components together, estimate the computing requirements
of each step (floating-point-operation count), and make a first preliminary
optimization of the performance of the search for a given available
computer power.  We show that the speedup of the
method is enough to make area searches practical with the kind of
computers that GEO600 may have available to it.

Probably more than $10^8$ neutron stars have been formed in the Galaxy in
its evolution.  Only about $10^3$ are known as pulsars or X-ray sources.
GEO600 and other gravitational wave projects plan to do directed
searches for radiation from those that are known to be spinning rapidly
enough for their radiation to be in the detection window (above 50~Hz for
GEO600).  But it would also be very interesting to identify new objects
by the gravitational waves they emit: if even a small fraction of
the unidentified neutron stars in the Galaxy are detectable, we would
learn much from them about stellar evolution and the physics of neutron
stars. To find such objects requires a blind search.

Recent observations and theoretical work give some reason to believe
that some stars may be detectable.  The recently discovered $r$-mode
instability\cite{Owen1998}
will lead to strong radiation from very young stars ($\sim 1$~yr),
and a residual radiation may persist for longer times.  The youngest
neutron star in the Galaxy may be only 40~y old, and if SN1987a contains
a neutron star then it will be very young indeed. There are suggestions
that low-mass X-ray binaries (LMXB's)are strong gravitational wave
emitters\cite{Bildsten1998}, and some of these will probably be targetted by
GEO600.  But there may be systems where the X-ray emission has turned
off relatively recently but the gravitational wave emission continues at
some level, and these could turn up in an area search.  And of course
there may be unexpected reasons for strong emission: the physics of
neutron stars is complex and not at all understood.

Another reason for being interested in developing search methods is
that they could be applied to other important problems.  Detecting
radiation from known LMXB's may require searches over large parameter
spaces, as almost certainly will the detection of radiation from
r-mode spindown after a neutron star has been formed in a supernova.
Radio searches for pulsars in binary systems have much in common with
our problem, and methods developed for gravitational waves could be
used in radio astronomy.

These algorithms are of particular interest to the GEO600 project because of
the likelihood that it will do a significant amount of observing as a
single detector, not in coincidence with other detectors.  According
to the published plans, GEO600 may be taking data of quality before
the large LIGO and VIRGO projects, and it will be unique among the first
interferometers in being able to run in narrow-band mode (signal
recycling), which will give it better sensitivity to continuous
signals than the larger projects would have in the selected bands.
Area surveys are likely to be among the first priorities for GEO600's
operations when they begin in 2000.

\section{Intrinsic difficulty of blind surveys}
The computational cost of a blind survey is large because the signal,
although predictable, depends on a number of parameters, and there is
a very large number of distinguishable sets of values for these
parameters.  This is primarily because detectors must observe
for times of order $10^7$~s in order to have reasonable sensitivity.
During this time the motion of the Earth produces very substantial
phase modulation in the received signal, and the spindown of
a neutron star can produce a significant change in its frequency.
Since signals can only be found if one can track their phase
accurately to within one cycle over the entire observing period,
one must search for signals over a huge parameter space.

Consider the effect of phase modulation.  The detector is carried
by the Earth as it orbits the Sun, and so if it observes a steady
source coherently for many months then it effectively synthesizes
a gravitational wave telescope with a diameter approaching 2~AU:
this is similar to the aperture synthesis common in radio astronomy.
The angular resolution of such a telescope is of order
$\lambda\/2{\rm {AU}}$, or about 0.2 arc seconds.  There are about
$10^{13}$ such resolution patches on the sky, each of which
impresses a distinguishable pattern of phase modulation onto the signal.

In addition there is spindown, which has been discussed extensively
elsewhere.\cite{Brady1998}  Young stars in particular can require
parametrization of the first three time-derivatives of the period,
and this can lead to large ($10^{10}$)
increases in the size of the parameter space.

To search this many parameter sets coherently (i.e. with the optimum
sensitivity that can be achieved by matched filtering),
each time treating the $2\times10^{10}$
data points that are sampled in 4~months for observations of up to
1~kHz, is beyond the capacity of any existing or planned computer.
In GEO600, the realistic available computing power that can be
dedicated exclusively to searches will be of the order of 20~Gflops.
Moreover, this computer is likely to be a loosely coupled set of parallel
processors (a cluster) rather than a tightly coupled parallel machine.
This means that the algorithm must require a minimum of inter-processor
communication.  Standard signal-processing techniques based on long
FFTs do not automatically satisfy this constraint.

\section{Hierarchical methods}
A solution to this mismatch is to use hierarchical procedures.  In
general, these involve a step in which candidate sources are selected
on the basis of a sub-optimal search, and then they are followed up
somehow to test whether they are real or just artifacts of noise.
The full data set is never searched at full sensitivity.

The initial selection of candidates inevitably runs the risk that
sources will be missed, but in some circumstances this risk is
smaller than one might expect.  In particular, the large parameter
space required for a blind area search implies that there will
be many opportunities for noise to masquerade as a signal, so
that confidence in detection will require a relatively high
signal-to-noise ratio even in optimum filtering.  It is conventional
in this problem to expect that the best one can hope for is an
amplitude SNR of 10.

If this is the case, then one can try to find a method in which
the initial sub-optimal is at a level such that a signal of this
strength would be likely to get through it.  Then very few
detectable signals will be lost in such a method.

Our proposed search method consists of the following stages for
treating a data set gathered in a total observing time of $T_{obs}$.
The data set is divided into shorter segments of length $T_c$, for
which a full coherent search is performed over the (much smaller)
parameter space appropriate to that length of data.  The power
spectra produced from each such period of time are searched for
evidence of a signal whose frequency is changing over the longer
time $T_{obs}$ in precisely the pattern expected for some one of the
parameter sets, using a method we have adapted from high-energy
experimental physics, called the Hough transform.\cite{gwda3}  And
finally, candidates are selected at the end of the Hough stage for
matched-filtering follow-up over the whole $T_{obs}$. This uses
approximate short-period Fourier transform techniques that we have
also developed, to avoid performing large FFTs on parallel
computers.\cite{Schutz1998}

For a given observing time $T_{obs}$, which we imagine is set
by operational considerations in the experiment, one is free
to choose the coherent search length $T_c$ as one wants.  The
longer it is, the more sensitive will be the search and the
more computer power will be required.  (Optimum
searches would use $T_c=T_{obs}$, but this make impractical
demands on computing.)  Changing $T_c$ affects the required
computing power of the different stages in different ways,
and this in turn will depend on the model of the signal: a
search will need to define reasonable parameter ranges.
We therefore present in this paper the
first attempts to {\em optimize} the search strategy, by
fixing the computer power available and choosing $T_c$ in
such a way as to give the best search sensitivity.

A different hierarchical approach to the same problem is being
developed in the LIGO project.\cite{Brady1999}  In this method,
the initial coherent transforms are combined using power-spectrum
summation, and the final follow-up is done with further power-spectrum
summation.  It is very important that these two different styles
be thoroughly evaluated to decide on their costs and returns for
different kinds of problems.

\section{Computational cost of the hierarchical procedure}
We describe the algorithm in some detail here.  It is useful to
establish our notation and basic concepts at the beginning.  We
aim at a total observing time $T_{obs}$ of order $10^7$~s, and
we expect to do the coherent searches over shorter time $T_c$ of
order 1~day.  In fact these searches are built from Fourier
transforms of even shorter data sets of length $T_s\sim 30$~min, in
a way we will describe.  At each stage in the calculation one must
consider the number of resolvable sets of parameters.  We call each
set a ``patch'' in parameter space, and let $N_p$ represent the
number of such patches needed in a calculation.  Longer data sets
need much larger values of $N_p$.  We search for sources in
a frequency bandwidth $B$, with a maximum frequency $f_0$.

The hierarchical procedure described above consists of three basic steps:

\begin{itemize}
  \item
I. {\it{matched filtering}} on chunks of data of duration $T_c$
much shorter than the total observation time.
\noindent
This stage is often referred to
as the {\it{coherent stage}},
since it employs the full information of the data, amplitude and phase.
It is computationally intensive for the reasons mentioned in the previous
section, but it is affordable because the time baseline is short.
The outcome of this is a set of $N_p$ FFTs for each data chunk. $N_p$ is
the number of patches in parameter space, and every FFT is demodulated
according to a patch. This means that if a signal from a patch
were present in the data, in the corresponding demodulated time series
it would appear as monochromatic and in the corresponding power
spectrum the power would be confined to one (two, at the most) frequency
bins. Signal to noise ratio in each chunk is $\sqrt{T_c\over T}$ times
smaller than it would be by matched filtering over $T$.
\noindent
Each filtered FFT of baseline $T_c$ is actually constructed from a set of
shorter baseline ($T_s$) FFTs, as mentioned in the previous section. The
short FFTs should belong to a frequency domain data base which should be
constructed as data is acquired. Such data base, as first suggested by
Frasca (\cite{frasca}), should be such that any periodic signal at a given
frequency in the data should appear as monochromatic during the observation
time baseline. This fixes the maximum length of the baseline at any given
frequency. In order to observe a range of frequencies one should ask that
the above requirement be fullfilled for the highest frequency of the range.
If $f_0$ is such frequency, then the time baseline for the FFTs of the data
base can be chosen to be $T_s=5.5\times10^3 \sqrt{\dps{300~{\rm{Hz}}\over
f_0}}$ s.  The use of the short-term frequency database means that
area searches can in practice be confined within specific narrow frequency
bands: all the data for a particular band can be loaded into the memory
of one processor, and negligible data transfers are needed between
processors.

   \item
II. the information from the each set of chunks is pieced
together by using the {\it{spectra}}
computed from the demodulated FFTs (\cite{lisa}, \cite{gwda3},
\cite{gwda4}).
\noindent
 The phase information is lost and
this is the reason this stage is said to be {\it{incoherent}}.
We can show, both analytically and with numerical experiments, that the
gain in signal to noise ratio - with respect to that of the
previous stage - is the factor $\sqrt[4]{T/T_c}$.
\noindent
At the end of this stage a threshold is set that selects suspect candidates
in the parameter space describing the signals one is searching.
\noindent
The strategy that we propose to use for this stage employs a technique which
is well known in image processing: the Hough transform (HT, hereafter). It
is called ``transform'' because it is indeed a transformation between the
data and the space of the parameters that describe the signal. The outcome
of the HT is an histogram in parameter space and significant clustering in a
pixel of parameter space indicates ``suspect'' consistency of data with a
signal
having the parameters of that pixel. Since the distribution of the number
count in each pixel can be known - in fact it is a Poisson distribution -
one can associate a false alarm probability, $p_{fa}$, to each pixel, which
measures the significance of the clustering obtained there. The threshold
$K$ is set on this quantity and defines the fraction of parameter space
around which a refined search will be performed by step III.

\item
III a coherent search, as described in step I, is repeated but with the
two following differences: 1) the baseline of the FFT is $T_{obs}$
2) the patches in parameters one corrects for, are the ones ``around'' the
candidates produced by step II. This stage is refered to as the ``follow
up'' stage.
\end{itemize}

The computational cost of each of these steps can be computed and thus the
total computational load may be equated to the available resources. This
defines the best acquirable sensitivity and the best choice of algorithm
parameters, as we shall show in the following.

Let $P$ indicate the total computation power available, $f_0$ and $B$
respectively the maximum intrinsic frequency and the band one wants to
search over, $N_{spin}$ the total number of sets of values of all
the spin-down parameters (which will differ for each step),
and $A$ the area of the sky (in steradians) where the search is confined to
(e.g. for an all-sky search $A=4\pi$).

The number of floating point operations $\chi$ necessary for each step is
the following:
\begin{eqnarray}
\chi_{I} & = & 1.2\times 10^{15}
\left(\dps{T_{obs}\over 10^7 ~{\rm{s}} } \right)
\left(\dps{f_0\over 300  ~{\rm{Hz}}} \right)^2
\left( \dps{B\over 300 ~{\rm{Hz}}}\right)
\left( \dps{T_c\over ~{\rm{1~ day}}}\right)^2
\left( \dps{A\over 4\pi}\right)
N_{spin,I}(T_c)\cr
\chi_{II}&=& 2.5\times 10^{16}
\left(\dps{T_{obs}\over 10^7 ~{\rm{s}} } \right)
\left(\dps{f_0\over 300  ~{\rm{Hz}}} \right)^2
\left( \dps{B\over 300 ~{\rm{Hz}}}\right)
\left( \dps{T_c\over ~{\rm{1~ day}}}\right)^2
\left( \dps{A\over 4\pi}\right)
N_{spin,II}(T_c,~T_{obs})\cr
 \chi_{III}&=& 4.5\times 10^{23}
\left(\dps{T_{obs}\over 10^7 ~{\rm{s}} } \right)^3
\left(\dps{f_0\over 300  ~{\rm{Hz}}} \right)^2
\left( \dps{B\over 300 ~{\rm{Hz}}}\right)
\left( \dps{A\over 4\pi}\right)
N_{spin,III}(T_{obs}) ~p_{fa}(K)
\end{eqnarray}
The cost for step I is the construction and filtering of FFTs over a
baseline $T_c$ starting from sets of baseline $T_s$.
\noindent
 The cost for step II is the cost of constructing the HT histograms in the
hyperspace of parameters.
\noindent
 The cost for step III is the cost of performing a coherent search over the
entire observation period around the suspect parameters identified by the
previous step. The area around the suspect parameter to be searched is the
resolution in parameter space at the end of the HT step (II).

\section{Optimization scheme}

For a given observation time and fixed computational power $P$, assuming
that data analysis should run at the same speed as data acquisition, the
maximum number of operations that one can perform is
$$\chi(T_c,K)=\dps{P T_{obs}}.$$
We can, though, choose how to distribute the computational load among steps
I, II and III. In particular we are free to choose the length of the time
baseline $T_c$ and the optimization scheme tells us how to do so in order to
acheive the maximum sensitivity. The point is the following: $P$ determines
the maximum number of follow-ups that one can perform, i.e. where one should
set the threshold $K$. On the other hand, for a signal of a given amplitude
one may determine, given $T_c$, what $p_{fa}$ it is expected to show up
with, after the HT procedure. Conversely, for any given $T_c$ and threshold
$K$, one can say what is the signal to noise ratio $\left( S\over
N\right)^2_{min}$ at the end of the follow-up stage for the signal which is
expected to exceed the threshold $50\%$ of the times. This is what we shall
refer to as the smallest expected detectable signal:
\be
p_{fa}(S/N)\sim e^{-{1\over 2}\sqrt{T_c\over T_{obs}}(S/N)^2}~\rightarrow ~
\left( {S\over N}\right)^2_{min}\sim 2 \sqrt{T_{obs}\over T_c} ~\ln {1\over
K}
\label{pfa}
\ee
For given computational power, the optimization scheme consists in
determining $T_c$ in such a way that the corresponding expected minimum
detectable signal is the smallest possible. Table \ref{t:opt} shows the
result of this optimization for the particular case of a search for
old (spindown age greater than $10^9$~y), fast ($f_0=1$~kHz) pulsars
using 4 months of data. With
20~Gflops of computing power, it would be possible to detect with
$50\%$ efficiency sources with SNR$\sim 23$, at the end of the follow up.
Performance unfortunately is a very slowly varying function of computing
power. In fact if the computational resources were increased by a factor
$50$ the corresponding $S/N_{min}$ would be reduced only by a factor of 2.
\begin{table}[h]
\label{t:opt}
\begin{center}
\begin{tabular}{|c|c|c|c|c|c|c|c|}
\hline \hline
$P$ & $T_{obs}$ & $f_0$ & $B$ & $N_{spin}$ & $A$ & ${T_c}_{/opt}$ &
${S/N}_{min}$ \\ \hline
$20$ Gflops & $10^7$ s & $1000$ Hz & $500$ Hz & $1$ & $4\pi$ & $14$ h &
$\sim 23.3 $\\
$100$ Gflops & $10^7$ s & $1000$ Hz & $500$ Hz & $1$ & $4\pi$ & $1.3$ days &
$\sim 18.2 $\\
$1000$ Gflops & $10^7$ s & $1000$ Hz & $500$ Hz & $1$ & $4\pi$ & $4$ days &
$\sim 12.8 $\\
$20$ Gflops & $10^7$ s & $1000$ Hz & $500$ Hz & $1$ & $0.6$ & $2.6$ days &
$\sim 14.6 $\\
$40$ Gflops & $10^7$ s & $1000$ Hz & $500$ Hz & $1$ & $0.6$ & $3.7$ days &
$\sim 13.1 $\\
$40$ Gflops & $10^7$ s & $1000$ Hz & $25$ Hz & $1$ & $4\pi$ & $3.6$ days &
$\sim 13.2 $\\
\hline\hline
\end{tabular}
\end{center}
\caption{For different computational resources,
and observation time $T_{obs}=4$ months, this table shows the performance
of the hierarchical algorithm presented in this pape, for searches for old
fast pulsars ($f_0=1$ kHz and no spindown parameters). The performance is
expressed as the signal to noise ratio, ${S/N}_{min}$, necessary for a
signal to be detect with a $50\%$ efficiency. Note that this signal to noise
ratio refers to a  coherent search over the entire $T_{obs}$.}
\end{table}

\section{Conclusions}
We have described a hierarchical, highly parallel algorithm to perform
wide area searches for continuous gravitational wave signals.  We have
shown that it can be optimized in such a way that, for an observing
period of $10^7$~s and a 20~Gflops computer,
a wide-band wide-area search for old,
high-frequency pulsars (typical of millisecond pulsars) can reach
an amplitude S/N limit of 23, which is a factor of 2.3 worse than
the best one could expect to do with matched filtering (considering
the confidence requirements mentioned earlier).  If the search is
over a more restricted area, such as the galactic plane, then the
sensitivity improves to about S/N of 15, only 50\% worse than optimum.
Restricting a search for millisecond pulsars to the galactic plane is
in fact very reasonable: they are formed in binary systems, so they
do not acquire the large space velocities of isolated pulsars, and
they should accumulate in the galactic plane forever.  Radio observations
can detect only nearby pulsars, perhaps less than 10\% of the galactic
population.

These numbers are very encouraging. We have not by any means exhausted
the possibilities for speeding up the algorithm, nor have we performed
the fullest possible optimization.  The final sensitivities can only be
better than those quoted here.




\section*{References}
\begin{thebibliography}{99}

\bibitem{SCHUTZ1991a}Schutz, B.F., ``Data Processing Analysis and Storage
for
Interferometric Antennas'', in Blair, D.G., ed., {\em The Detection of
Gravitational Waves}, (Cambridge University Press, Cambridge England, 1991),
406--452.

\bibitem{Brady1998}Brady, P.R., Creighton, T., Cutler, C. and Schutz, B.F.
``Searching for periodic sources with LIGO'', {\em Phys.Rev.} {\bf D57},
2101-2116 (1998).

\bibitem{Schutz1998}Schutz, B.F., ``Sources of radiation from neutron
stars'', in {\em Second Workshop on Gravitational Wave Data Analysis}, ed.s
M.
Davier and P. Hello (Editions Frontieres 1998).

\bibitem{gwda3}Papa, M.A., Astone, P., Frasca, S. and Schutz, B.F.,
{\it ``Searching for continuous waves by line
identification''} in {\em Second Workshop on Gravitational Wave
Data Analysis}, eds. M.
Davier and P. Hello (Editions Frontieres 1998).

\bibitem{lisa} Papa, M.A., Schutz, B.F., Frasca, S.,
{\it{ ``Detection of Continuous Gravitational Wave Signals:
Pattern Tracking with the Hough Transform''}} in {\em International LISA
Symposium on the detection and Observation of Gravitational Waves in Space},
ed. W.M. Folkner (AIP Conf. Proc., 1998).

\bibitem{Owen1998}Owen, B., Lindblom, L, Cutler, C., Schutz, B.F., Vecchio,
A., Andersson, N., ``Gravitational waves from hot young rapidly rotating
neutron stars'', gr-qc/9804044.

\bibitem{Bildsten1998}Bildsten, L., ``Gravitational Radiation and Rotation
of Accreting Neutron Stars'', astro-ph/9804325.

\bibitem{Brady1999}Brady, P.R., Creighton, T. `` Searching for periodic
sources with LIGO. II: Hierarchical Searches'', gr-qc/9812014
and private communication.

\bibitem{frasca} Proc. of Aspen Winter Conf. on Gravitational Wave
Detection, January 1997, Aspen

\bibitem{gwda4}
http://fiji.nirvana.phys.psu.edu/gwdaw98/Proceedings/Papa/index.html
\end{thebibliography}
\end{document}